\documentclass[%
 reprint,
showkeys, 
amsmath,amssymb,
aps,
]{revtex4-1}

\usepackage{graphicx}
\usepackage{dcolumn}
\usepackage{bm}
\usepackage{circuitikz} 
\usepackage{natbib}
\usepackage{mathtools}
\usepackage{siunitx}
\usepackage{bbold}
\usepackage{braket}
\usepackage{bm}
\usepackage{pgfplots}
\usepackage{pgfplotstable}
\usepgfplotslibrary{statistics}
\usepackage{hyperref}
\pgfplotsset{
compat=1.8,
cycle list={blue\\red\\},
boxplot/hide outliers/.code={
	\def\pgfplotsplothandlerboxplot@outlier{}%
},
}
\pgfmathdeclarefunction{fpumod}{2}{%
    \pgfmathfloatdivide{#1}{#2}%
    \pgfmathfloatint{\pgfmathresult}%
    \pgfmathfloatmultiply{\pgfmathresult}{#2}%
    \pgfmathfloatsubtract{#1}{\pgfmathresult}%
    \pgfmathfloatifapproxequalrel{\pgfmathresult}{#2}{\def\pgfmathresult{3}}{}%
}

\begin{document}

\preprint{APS/123-QED}

\title{A quantum alternating operator ansatz with \\ hard and soft constraints for lattice protein folding}
%
\author{Mark Fingerhuth\footnote{Corresponding author.}}%
 \email{All correspondence to mark@proteinqure.com}
 \affiliation{ProteinQure Inc., Toronto, Canada}
 \affiliation{University of KwaZulu-Natal, Durban, South Africa}
\author{Tomas Babej}
 \affiliation{ProteinQure Inc., Toronto, Canada}
\author{Christopher Ing}%
 \affiliation{ProteinQure Inc., Toronto, Canada}
\date{\today}

\begin{abstract}
Gate-based universal quantum computers form a rapidly evolving field of quantum computing hardware technology.
In previous work \citep{pq_whitepaper1}, we presented a quantum algorithm for lattice protein folding on a cubic lattice, tailored for quantum annealers.
In this paper, we introduce a novel approach for solving the lattice protein folding problem on universal gate-based quantum computing architectures.
Lattice protein models are coarse-grained representations of proteins that have been used extensively over the past thirty years to examine the principles of protein folding and design.
These models can be used to explore a vast number of possible protein conformations and to infer structural properties of more complex atomistic protein structures.
We formulate the problem as a quantum alternating operator ansatz, a member of the wider class of variational quantum/classical hybrid algorithms.
To increase the probability of sampling the ground state, we propose splitting the optimization problem into hard and soft constraints.
This enables us to use a previously underutilised component of the variational algorithm to constrain the search to the subspace of solutions that satisfy the hard constraints.
\end{abstract}

\maketitle


\section{Introduction}
\label{sec:introduction}

Proteins are molecular machines which serve a vast number of functions within living organisms.
Each protein consists of a chain of amino acids which can spontaneously fold into three-dimensional shapes that defines its function.
For most natural proteins, three-dimensional structure cannot be predicted from the amino-acid sequence alone due to the large space of available folds \cite{dill2008the}.
This limitation results in a significant number of clinically relevant proteins having unknown structures \cite{edwards2009large}, preventing the rational discovery of treatments for devastating diseases like cancer, cardiovascular and neurodegenerative diseases.
Over the past several decades, computational modeling has served as an invaluable tool for studying protein structure and function.
In this work we introduce a new algorithmic approach for the computational modeling of proteins on quantum computing devices that may offer significant improvements as these devices scale in computing capability.

Numerous computational methods exist for the determination of protein structure from an arbitrary input sequence, although all of these techniques have significant computational limitations \cite{kmiecik2016coarse}.
High-resolution atomistic structures can be determined using both physics-based or knowledge-based methods, but both these methods are fundamentally restricted in the space of structures they can search, due to either the timescale of simulations required for folding or the size of the protein structure database \cite{dill2012the}.
Coarse-grained lattice models have the advantage that they can be utilized to enumerate and score a vast number of candidate protein folds, which can be useful for analyzing general sequence-structure properties and verifying analytical models \cite{mirny2001protein, agozzino2018protein}.
In the hydrophobic-polar (HP) model, all amino acids in a sequence are classified as either hydrophobic (H) or polar (P), enabling the study of the fundamental energetic principles of folding \cite{dill1985theory, lau1989lattice}.
In the Miyazawa-Jernigan (MJ) model, all twenty natural amino acids are uniquely represented, each of which has distinct pairwise interactions parameterized using known structural data \cite{miyazawa1985estimation,miyazawa1996residue}.
Both of these models involve the representation of protein chains as a self-avoiding walk on a grid, which can vary in geometric complexity (planar, cubic, hexagonal).
The energy of a fold is determined by the sum of pairwise interactions between amino acids as determined by a look-up table, and the native state of the protein is the configuration with the lowest energy.
Despite the simplicity of the model, finding the minimum energy lattice fold is an NP-complete problem \cite{hart1997robust,unger1993finding,berger1998protein}.
As such, new computational methods need to be developed to determine the minimum energy lattice fold of larger length proteins (greater than 100 amino acids).

\begin{figure}[b]
	\includegraphics[scale=0.35]{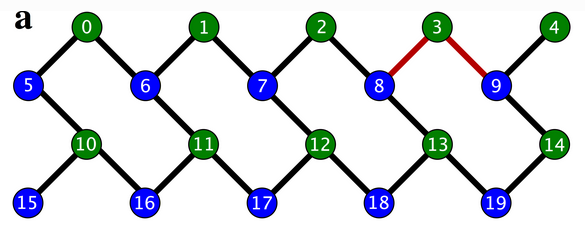}
\caption{\label{img:acorn_hardware_graph}Visualization of the planar lattice layout of the 19 qubit Acorn quantum processor by Rigetti Computing}
\end{figure}

\begin{figure*}
\includegraphics[scale=0.34]{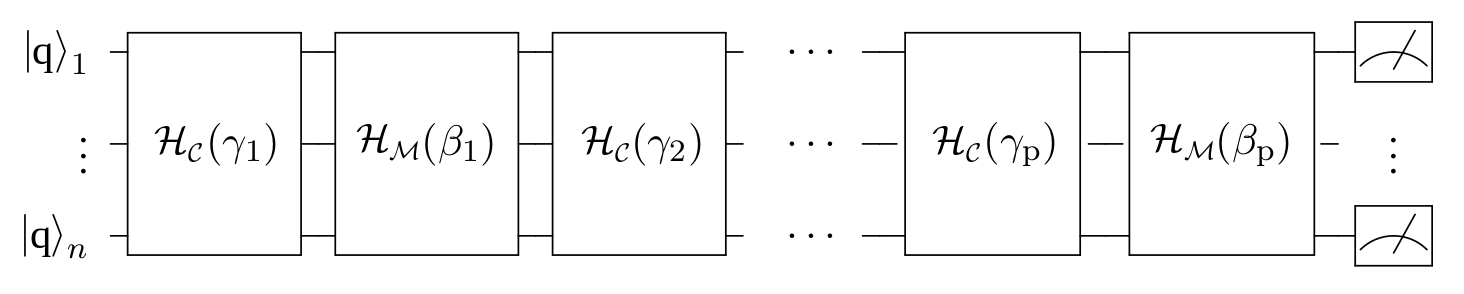}
\caption{\label{img:qaoa_circuit}A quantum alternating operator ansatz circuit of depth $p$.
The quantum circuit alternates between implementing the cost Hamiltonian $H_C$ parametrized by parameters $\gamma_i$ and the mixer Hamiltonian $H_M$ parametrized by parameters $\beta_i$.
The minimization algorithm on the classical computer then optimizes the parameters $\gamma_i$ and $\beta_i$ based on the measured outcomes.
This effectively maximizes the expectation value $\braket{\boldsymbol{\gamma, \beta} |\, C \,|\\ \boldsymbol{\gamma,\beta }}$ where $C$ is the cost function of the optimization problem.}
\end{figure*}

In past years, superconducting implementations of gate-based quantum computers have scaled drastically with Google's \textit{Bristlecone} processor leading the newest generation of chips with more than 50 qubits and the capacity of running quantum algorithms with a quantum circuit depth of up to 40 \texttt{CNOT} operations \citep{google}.
In almost all current hardware implementations, the qubits are arranged in a planar lattice structure as illustrated in the example in Fig.~\ref{img:acorn_hardware_graph} which shows the topology of Rigetti's \textit{Acorn} quantum processing unit with 19 qubits.
Even though the planar lattice layout allows for the implementation of the surface code, a promising quantum error correction protocol, it limits the connectivity of individual qubits to nearest neighbour interactions \citep{surfacecode1,surfacecode2}.
Due to the immense resource requirements of quantum error correction, most likely none of these qubits will be fully error corrected in the near future. For this reason, these near-term quantum devices are referred to as Noisy Intermediate-Scale Quantum (NISQ) technology \citep{preskill2018quantum}.
Despite the presence of errors, it is widely believed that NISQ computers will be able to outperform classical computers in certain tasks and achieve 'quantum computational supremacy' or 'quantum advantage' in the near future \citep{google}.
One of the main reasons behind this believe are recent developments in a family of quantum algorithms known as \textit{variational algorithms}. These algorithms combine quantum and classical computers in a hybrid scheme in which the classical computer optimizes the parameters of a quantum circuit. They have the potential to demonstrate quantum computational supremacy due to algorithm design decisions specifically aimed to compensate for some of the disadvantages of NISQ computers \citep{farhi2016quantum}. As a result, a lot of recent attention has been focused on applying variational quantum algorithms to supervised classification, neural network training, prime factorization as well as quantum chemistry \citep{schuld2018circuit,ryabinkin2018qubit,verdon2017quantum,kandala2017hardware,anschuetz2018variational}.

This work will establish a framework for finding ground states of lattice proteins by means of variational quantum algorithms on universal gate-based quantum computers.
In order to use a variational algorithm, the problem first needs to be encoded as a Hamiltonian, whose ground state corresponds to the correct lattice fold of the protein.
\citet{perdomo2008construction} first developed a Hamiltonian encoding by imposing an absolute coordinate system onto the lattice.
In subsequent work, \citet{babbush2012construction} devised three novel encodings, one of which was used by \citet{perdomo2012finding} to establish a proof-of-concept by folding the 6-amino-acid protein PSVKMA on the quantum annealing processor D-Wave One. In previous work, we have generalized their Hamiltonian encodings from two-dimensional to three-dimensional lattices whilst decreasing circuit complexity from quadratic to quasilinear \citep{pq_whitepaper1}.
In this paper we report a novel encoding of the lattice protein folding problem using one-hot encoded turns designed for universal quantum computing architectures. This encoding produces Hamiltonians that can be minimized with quantum alternating operator ansatz circuits, while enabling the usage of several different mixer Hamiltonians that enforce varying subsets of optimization constraints.


\subsection{Quantum/classical variational algorithms}

Variational algorithms are an active field of research due to their promising applicability during the NISQ era.
These hybrid algorithms consist of a fixed-length sequence of parametrizable quantum gates that are executed on a quantum computer whilst the gate parameters are \textit{variationally} optimized by a classical computer.
An important property of this approach is that it is robust against systematic errors on the quantum computer since the classical optimization step of the algorithm can correct for them in parameter space \citep{farhi2017quantum}.
Furthermore, their requirements for quantum circuit depth (and therefore qubit coherence time) are highly flexible, which makes them particularly attractive for NISQ era quantum computers.
The two cornerstones of variational algorithms are the quantum approximate optimization algorithm by \citet{farhi2014quantum} and the variational quantum eigensolver by \citet{peruzzo2014variational}.
The quantum approximate optimization algorithm was later generalized to the quantum alternating operator ansatz by \citet{hadfield2017quantum} which we will refer to as QAOA.

QAOA circuits can be seen as trotterized quantum annealing where the order of trotterization $p$ determines the required quantum circuit depth and quality of the resulting solutions.
Note that in the limit of $p\rightarrow\infty$ the QAOA circuits are equivalent to adiabiatic quantum computation \citep{farhi2014quantum}.
QAOA-type algorithms sequentially apply alternating circuits to the initial quantum state; the two parametrized circuits implementing the cost Hamiltonian $H_{C}$ and the mixer (or driver) Hamiltonian $H_{M}$.
Fig.~\ref{img:qaoa_circuit} shows a QAOA circuit of depth $p$ whereby the $i$-th application of $H_C$ and $H_M$ is parametrized by parameter $\gamma_i$ and $\beta_i$ respectively.
Let us denote the resulting quantum state $\ket{\boldsymbol{\gamma,\beta}}$ where $\boldsymbol{\gamma}=\{\gamma_1, ..., \gamma_p\}$ and $\boldsymbol{\beta}=\{\beta_1, ..., \beta_p\}$.
A classical minimization algorithm such as Nelder-Mead \citep{nelder1965simplex} is then used to optimize the parameters $\gamma_i$ and $\beta_i$ on a classical computer such that the expectation value,
\begin{equation}
	\label{equ:expect_value}
	\braket{\boldsymbol{\gamma, \beta} |\, C \,|\\ \boldsymbol{\gamma,\beta }}\, ,
\end{equation}
is maximized.
Traditionally, the cost Hamiltonian encodes the entire cost function representing the optimization problem at hand.
$H_{C}$ separates the possible solutions by phase and consists only of $Z$ terms. It follows that $H_{C}$ is a purely classical Hamiltonian.
The mixer Hamiltonian $H_{M}$ is 'mixing' the amplitudes between different solution states and can be regarded as the QAOA component exploring the solution space.
To mix amplitudes between states with different eigenvalues, $H_{M}$ should be chosen in a way such that it does not commute with $H_{C}$, that is $[H_{C}, H_{M}] \neq 0$.
In the original work by \citet{farhi2014quantum} a pure X mixer of the form
\begin{equation}
	H_{M} = \sum_{i} X_{i},
\end{equation}
where the sum is over all qubits, was chosen.
However, in QAOA implementations on universal quantum computing architectures we can freely choose $H_{M}$ depending on the problem at hand.
This stands in contrast with current quantum annealing hardware, where the mixer Hamiltonian is fixed by the chosen hardware implementation and cannot be defined by the user \citep{hadfield2017quantum}.
Recent work by \citet{hadfield2017quantum} has shown that a custom mixer Hamiltonian can be used to encode parts of the optimization problem.
More specifically, if the constraints of the optimization problem can be broken up into hard and soft constraints (defined below) the mixer Hamiltonian can be used to encode the hard constraints whilst the cost Hamiltonian encodes the remaining soft constraints.
After measuring the resulting quantum state, the solutions to the encoded optimization problem are represented as $n$-bit bitstrings.

An optimization problem can be defined as the problem of finding the best solution from all feasible (or candidate) solutions.
However, based on the properties of the problem's encoding not every possible bitstring $\mathbf{x} \in \{0,1\}^n$ represents a feasible solution.
In other words, the space of all feasible solutions to the given optimization problem projected under the encoding often forms a proper subspace of $\{0,1\}^n$.
The size of the space of all unfeasible solutions varies depending on the constraints used.
Yet, speedups can be often obtained by modifying the optimization algorithm as to avoid the space of unfeasible solutions.

\textit{Hard constraints} are conditions that must be satisfied for a solution to be a \textit{feasible} solution e.g. "every solution string must have a Hamming weight of three".
We  define the \textit{feasible subspace} as the space of all feasible solutions.
\textit{Soft constraints} are additional constraints in the optimization problem that do not necessarily need to be satisfied for a solution to be feasible e.g. "low energy solutions require the three $1$'s in the bitstring to be consecutive".
Note, that optimal solutions to the optimization problem satisfy both hard and soft constraints.

\section{One-hot encoded lattice proteins}
\label{sec:onehot_encoding}

This section formulates the lattice protein folding problem in a way that allows the separation of optimization constraints into hard and soft constraints.
The proposed encoding requires $6N-17 \in O(N)$ qubits in order to encode a lattice protein with $N$ amino acids.
We will first discuss the case of protein folding on cubic lattices and subsequently simplify our encoding to planar lattices with near-term implementation in mind.

\subsection{Cubic lattices}
\label{sec:cubic_lattices}
\begin{figure}[t!]
	\includegraphics[scale=0.6]{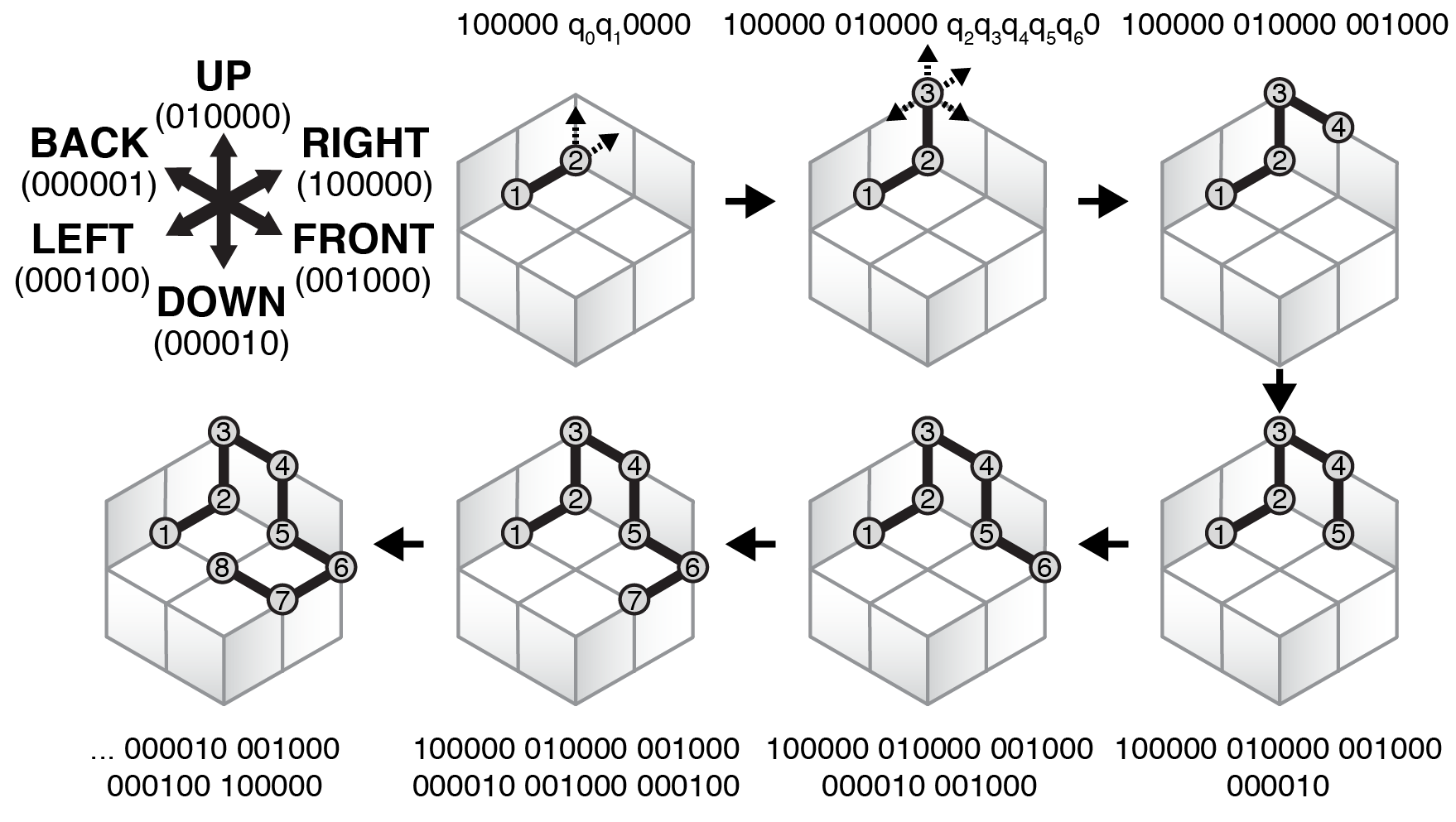}
\caption{\label{img_onehot_directions}Visualization of the six one-hot encoded spatial directions for cubic lattice protein folding. The sequence of images shows an example of the encoding using a low-energy cubic lattice fold for a protein with 9 residues. Note, that due to mirror and rotational symmetries we can always fix the zeroth turn to $100000$, the second turn to $q_0q_10000$, which only allows \texttt{RIGHT} or \texttt{UP}, and finally the second turn to $q_2q_3q_4q_5q_60$ such that it cannot go to the \texttt{BACK}.}
\end{figure}

In our previous work \citep{pq_whitepaper1}, we introduced two turn-based encodings of the lattice protein folding problem on a cubic lattice that use dense, three-qubit representations of spatial directions on the lattice.
Turn-based encodings of the lattice protein folding problem represent the lattice fold as a sequence of moves that constitutes a walk on the lattice, rather than encoding the absolute lattice coordinates of each individual amino acid.
In this work, we propose a sparser turn-based encoding that uses $6$ qubits to represent each of the six spatial directions.
One-hot encoding data is a commonly used preprocessing technique in machine learning where each element of a discrete set of $N$ elements is encoded as a bitstring of length $N$ with a Hamming weight of 1.
Fig.~\ref{img_onehot_directions} illustrates the one-hot encoding that is used in the remaining parts of this paper.
Due to rotational and mirror symmetries of the cubic lattice, we can, without loss of generality, fix $11$ qubits in the prefix of the solution string.
First of all, the choice of the direction of the first turn is arbitrary, as the lattice fold can be rotated on the lattice with no change to its energy (since the set of nearest neighbour contact pairs remains the same).
Hence the entire zeroth turn can be fixed - in our derivations, we chose the \texttt{RIGHT} ($100000$) direction.
The only two valid situations that can be distinguished under rotation for the next move are either a move in the same direction, or a perpendicular move, which corresponds to \texttt{RIGHT} and \texttt{UP} moves in our formulation. This allows us to fix $4$ qubits in the first move.
The second move can exploit the last remaining degree of freedom - mirror symmetry with respect to the plane defined by the \texttt{RIGHT} and \texttt{UP} move. In this situation, the \texttt{FRONT} and \texttt{BACK} moves are mirror-symmetric and as such we can prohibit one of them. We prohibit the move to the \texttt{BACK} by fixing its qubit in the sextuple to 0.

An obvious advantage of using one-hot encoded directions is that there exists a straightforward algorithm to determine whether a particular turn went into a particular direction.
Namely, within each sextuple that represents one turn, each possible direction uses exactly one qubit which serves as a binary flag indicating that this turn went its way.
For every qubit sextuple that encodes a turn $t$ we can thus define the mapping,
\begin{align}
	\label{eq:k_mapping1}
	\text{right} \rightarrow k=0, \\
	\text{up} \rightarrow k=1, \\
	\text{front} \rightarrow k=2, \\
	\text{left} \rightarrow k=3, \\
	\text{down} \rightarrow k=4, \\
	\label{eq:k_mapping5}
	\text{back} \rightarrow k=5,
\end{align}
where for each of the six directions $k$ is a pointer to the qubit (binary flag) that indicates whether turn $t$ went in this direction. For the remainder of this paper we will use the shorthand notation $q_{t,k} = q_{6t+k-11}$. 
Furthermore, we use notation $\bar{q}$ to denote a qubit that represents the opposite direction as the one represented by qubit $q$ within the same move.
For example, $\overline{q_{4, 1}}=q_{4, 4}$.

As previously mentioned, a one-hot encoded turn has a Hamming weight of 1.
Hamming weight is defined as the number of bits in the string that are equal to 1 or, equivalently, the Hamming distance between the bitstring and the all-zero bitstring of the same length.
Therefore, each turn will increment the total expected Hamming weight of the solution string by 1.

\vspace{3 mm}
\textbf{Construction of the mixer Hamiltonian $H_{M}$.}
We define the hard constraint that for every turn $t$ that is represented by $n$ bits all $n$ bits must sum to one in order for the string to represent a one-hot encoded direction. Hence,
\begin{equation}
\label{eq:hard_constraint1}
\sum^{n-1}_{b=0} q_{t,b} = 1\, .
\end{equation}
From this we can derive the more general second hard constraint that for every protein of size $N$ all $N-1$ turns must only contain a single one. Thus, a valid solution string for a protein of size $N$ must have a Hamming weight of $N-2$ (since the first fixed turn is not represented in the solution string):
\begin{equation}
\label{eq:hard_constraint2}
\sum^{N-2}_{t=1}\sum^{n-1}_{b=0} q_{t,b} = N-2\, .
\end{equation}
These two statements determine the feasible subspace of our problem and allow us to define mixer Hamiltonians which ensure that the search stays within the feasible subspace. In the next sections we will define two distinct types of mixer Hamiltonians, referred to as XZ and XY mixers, which we will ultimately benchmark in Section ~\ref{sec:results_discussion}.

\vspace{3 mm}
\textbf{Simple XY mixer}
Ideally, the mixer Hamiltonian should preserve Hamming weight and only move the $1$ within each sextuplet encoding a turn. To achieve this, the $\texttt{SWAP}_{i,j}$ operation is an ideal building block since it preserves Hamming weight. $\texttt{SWAP}_{i,j}$ maps $\ket{00}\rightarrow\ket{00}$, $\ket{01}\rightarrow\ket{10}$, $\ket{10}\rightarrow\ket{01}$ and $\ket{11}\rightarrow\ket{11}$.
\citet{hen2016quantum} define the generalized $\texttt{SWAP}_{i,j}$ as
\begin{equation}
\label{eq:trad_swap}
\texttt{SWAP}_{i,j} = \frac{1}{2} (\mathbb{1} + X_iX_j + Y_iY_j  + Z_iZ_j).
\end{equation}
The mixer Hamiltonian should not commute with the cost Hamiltonian (which consists purely of $Z$ terms) but in the current definition of the \texttt{SWAP} operation, the identity operation as well as the $Z_iZ_j$ component both commute with the cost Hamiltonian.
Thus, we drop these components and redefine $\texttt{SWAP}_{i,j}$ as
\begin{equation}
\label{eq:our_swap}
\texttt{SWAP}_{i,j} = \frac{1}{2} (X_iX_j + Y_iY_j).
\end{equation}
For a more rigorous analysis of the properties of this $\texttt{SWAP}$ operation the interested reader is referred to \citep{hadfield2017quantum}. The mixer Hamiltonian should sample from the distribution over all possible combinations of turns and, thus, we need to apply a \texttt{SWAP} operation to each pair of qubits within a turn $t$:
\begin{equation}
\label{eq:XY_Mt}
\text{M}_{t} = \sum^{n-2}_{i=0} \sum^{n-1}_{j=i+1} \texttt{SWAP}_{i,j}.
\end{equation}
By simply summing over all $N-2$ turns for a protein with $N$ amino acids we get the final XY mixer Hamiltonian,
\begin{equation}
\label{eq:xy_simple}
\text{H}_{M} = \text{XY}_{simple} = \sum^{N-2}_{t=1} \text{M}_{t}.
\end{equation}
Given that the initial state is a feasible bitstring or a superposition of such, XY$_{simple}$ ensures that the search only takes place in the feasible subspace.

\vspace{3 mm}
\textbf{Simple XZ mixer}
Instead of swapping the values of two qubits within an $n$-tuple we can flip individual qubits conditioned on the fact that all other qubits in the $n$-tuplet are $0$.
For this purpose, we define the conditional bit flip operator,
\begin{equation}
\label{eq:flip_operator_with_q}
	\text{F}_{t,k} = X_{t,k} \prod_{\substack{k_o \in [0,n]\\ k_o \neq k}} \bar{q}_{t,k_o}.
\end{equation}
Substituting $\bar{q}_{t,k_o} = \frac{1}{2} (1+Z_{t,k_o})$ yields
\begin{equation}
\label{eq:flip_operator_with_Z}
	\text{F}_{t,k} = X_{t,k} \prod_{\substack{k_o \in [0,n]\\ k_o \neq k}} \frac{1}{2} (1 + Z_{t,k_o}).
\end{equation}
Using this operator, the Hamming weight of the total bitstring does not remain constant as in the XY mixer. However, it can still be bounded since any $n$-tuplet can now either have a Hamming weight of $0$ or $1$. Hence, we need to adjust Eq.~\ref{eq:hard_constraint1} to
\begin{equation}
\label{eq:hard_constraint1_new}
\sum^{n-1}_{b=0} q_{t,b} =
\begin{cases}
0 \\
1
\end{cases}.
\end{equation}
As a result, the sum over the entire solution string in Eq.~\ref{eq:hard_constraint2} must now satisfy
\begin{equation}
\label{eq:hard_constraint2_new}
\sum^{N-2}_{t=1}\sum^{n-1}_{b=0} q_{t,b} \leq N-2.
\end{equation}
To allow for transitions between all feasible bitstrings we take the sum over
all bits within a given turn $t$:
\begin{equation}
\label{eq:XY_Mt}
	\text{M}_{t} = \sum^{n-1}_{k=0} \text{F}_{t,k}.
\end{equation}
Similar to Eq.~\ref{eq:xy_simple} we obtain the final XZ mixer Hamiltonian by summing over all turns
\begin{equation}
\label{eq:xz_simple}
\text{H}_{M} = \text{XZ}_{simple} = \sum^{N-2}_{t=1} \text{M}_{t}.
\end{equation}
Given the adapted notion of the feasible subspace for XZ mixers and an initial state that is either a feasible bitstring or a superposition of such, XZ$_{simple}$ only allows for state transitions within the feasible subspace.


\vspace{3 mm}
\textbf{Construction of the cost Hamiltonian $H_{C}$.}
The cost Hamiltonian encodes the objective function of our problem.
In lattice protein folding, the goal is to find minimum energy self-avoiding walks on the lattice. The energy of a given protein fold is computed as the sum of interaction energies of non-bonded nearest neighbour amino acids.
We enforce both objectives in the final cost Hamiltonian: to avoid self intersections we will introduce the component $H_{overlap}$, and to account for the interaction between amino acids we introduce another component, $H_{pair}$.
Hence, the total cost Hamiltonian is defined by,
\begin{equation}
\label{eq:}
	\text{H}_{C} = H_{overlap} + H_{pair}.
\end{equation}
In previous work \citep{pq_whitepaper1}, we derived both components for a more dense bit encoding of the directions.
However, to construct the components $H_{overlap}$ and $H_{pair}$ we had to introduce a boolean expression for each spatial direction, \texttt{RIGHT} ($+x$), \texttt{LEFT} ($-x$), \texttt{UP} ($+y$), \texttt{DOWN} ($-y$), \texttt{BACK} ($-z$) and \texttt{FRONT} ($+z$), that yields TRUE if a particular turn went into that direction.
The new one-hot encoding simplifies this step since each qubit serves as a binary flag directly indicating if a particular turn went into a particular direction.
Eq.~\ref{eq:k_mapping1} - \ref{eq:k_mapping5} already introduced a mapping between qubit indices within a turn $t$ and the six spatial directions.
Thus, to determine if turn $t$ went into direction $k$ we define the simple function:
\begin{equation}
	\label{eq:binary_flags3d}
	d^t_{k} = \left\{
	\begin{array}{ll}
		q_k  & \mbox{if } t=1 \mbox{ and } k \in \{0,1\}, \\
		0  & \mbox{if } t=1 \mbox{ and } k \notin \{0,1\}, \\
        0  & \mbox{if } t=2 \mbox{ and } k=5, \\
		q_{6t+k-10} & \mbox{otherwise.}
	\end{array}
\right.
\end{equation}
For the new one-hot encoding introduced in this paper, we will use the same 'sum string' approach to generate $H_{overlap}$ as in \citet{pq_whitepaper1}.
A sum string $s$ contains the number of turns the lattice protein has taken in the $k$ direction between any two residues $i$ and $j$ (for a detailed description see \citep{pq_whitepaper1}).
Given this, we can define:
\begin{equation}
	s^r_{k}(i,j) = r^{th}\,\text{digit of} \sum^{j-1}_{p=i} d^p_{k},
\end{equation}
The purpose of $H_{overlap}$ is to determine if any two residues overlap and subsequently penalize the overlap by increasing the energy for this particular bitstring. As in \citep{pq_whitepaper1} we define,
\begin{equation}
	\label{eq:h_olap}
	H_\text{olap} = \lambda_\text{olap} \sum^{N-3}_{i=0} \sum^{\lfloor\frac{(N-i-1)}{2}\rfloor}_{j=1} H_\text{olap}(i,i+2j) .
\end{equation}
where for $D$ spatial dimensions,
\begin{equation}
	\label{equ:h_olap_ij}
	H_\text{olap}(i,j) = \prod^D_{k=1} \Big( \prod^{\lceil\log_2(j-i)\rceil}_{r=1}\text{XNOR}(s^r_{k}(i,j),s^r_{\bar{k}}(i,j)) \Big).
\end{equation}
and $\text{XNOR}(p,q) = 1-p-q+2pq$. Note, that $\bar{k}$ was previously defined as the direction opposite to $k$.

In order to account for the interaction between two non-bonded nearest neighbour residues $i$ and $j$ we need to determine if they are nearest-neighbours, that is, if their lattice distance is equal to $1$.
For this reason, we introduce the same adjacency function as in \citep{pq_whitepaper1}:
\begin{align}
	a_k(i,j) = &\Bigg[ \prod_{w \neq k} \Big( \prod^{\lceil \log_2(j-i)\rceil}_{r=1} \text{XNOR}(s^r_{w}(i,j),s^r_{\bar{w}}(i,j)) \Big) \Bigg] \nonumber \\
			   &*\Bigg[ \text{XOR} (s^1_{k}(i,j),s^1_{\bar{k}}(i,j)) \nonumber \\
			   &*\prod^{\lceil \log_2(j-i)\rceil}_{r=2} \text{XNOR} (s^r_{k}(i,j),s^r_{\bar{k}}(i,j)) \nonumber \\
			   &+ \sum^{\lceil \log_2(j-i)\rceil}_{p=2} \Big( \text{XOR} (s^{p-1}_{k}(i,j),s^p_{k}(i,j)) \nonumber \\
			   &*\prod^{p-2}_{r=1} \text{XNOR} (s^r_{k}(i,j),s^{r+1}_{k}(i,j)) \nonumber \\
			   &*\prod^{p}_{r=1} \text{XOR} (s^r_{k}(i,j),s^{r}_{\bar{k}}(i,j)) \nonumber \\
			   &*\prod^{\lceil \log_2(j-i)\rceil}_{r=p+1} \text{XNOR} (s^r_{k}(i,j),s^{r}_{\bar{k}}(i,j)) \Big) \Bigg].
\end{align}
From this we can define $H_{pair}$ in $D$ dimensions as,
\begin{equation}
	\label{eq:h_pair}
	H_\text{pair} = \sum^{n-3}_{i=1}  \sum^{(N-i-1)/2}_{j=1} P_{i,1+i+2j}\sum^{D}_{k=1} a_k(i,1+i+2j).
\end{equation}
where $P_{i,1+i+2j}$ is the interaction strength between the $i$-th and the $(1+i+2j)$-th residue defined by the interaction matrix of either the HP or the MJ model.

\subsection{Planar lattices}
\label{sec:planar_lattices}
\begin{figure}
	\includegraphics[scale=0.57]{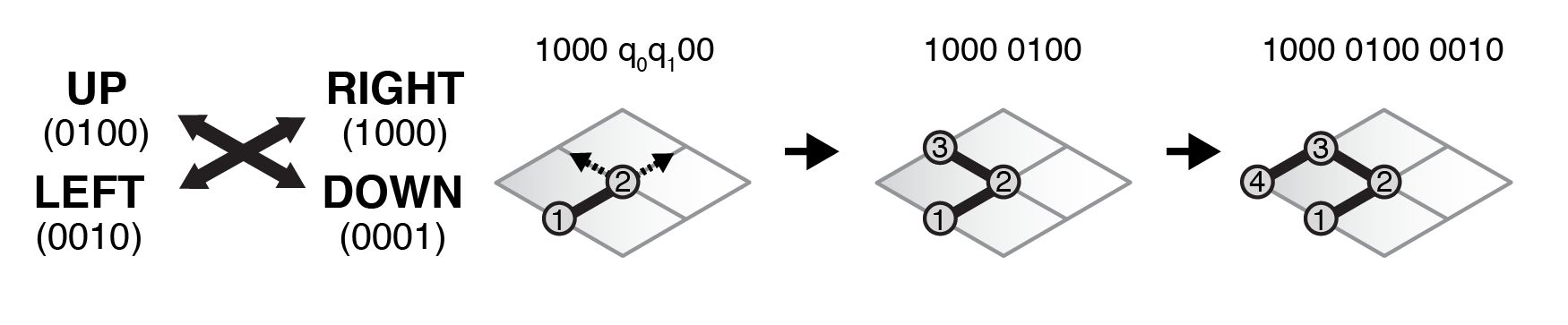}
\caption{\label{img_onehot_directions_2d}Visualization of the four one-hot encoded spatial directions for planar lattice protein folding. The figure also shows an example of a low-energy lattice fold on a planar lattice and its corresponding bitstring.}
\end{figure}

Considering the limitations of current quantum computing hardware, we focus on planar rather than cubic lattice proteins in the experimental implementation of this work.
On a planar lattice we only deal with four spatial directions and hence each direction requires only four qubits to be one-hot encoded.
The one-hot encoding of a protein of size $N$ on a planar lattice requires a total of $4N-10 \in O(N)$ qubits.
Fig.~\ref{img_onehot_directions_2d} shows the chosen one-hot encoding for the planar lattice including the fact that we fix the first turn to $1000$ (\texttt{RIGHT}).
Furthermore, the second turn consists of only two rather than four qubits since we can fix it to go either to the \texttt{RIGHT} or \texttt{UP} due to rotational symmetry.

The mapping between qubit indices and directions within a turn $t$ is given as right$\rightarrow k=0$, up$\rightarrow k=1$, left$\rightarrow k=2$ and down$\rightarrow k=3$.
From this the analogue of Eq.~\ref{eq:binary_flags3d} follows immediately:
\begin{equation}
	\label{eq:binary_flags2d}
	d^t_{k} = \left\{
	\begin{array}{ll}
		q_k  & \mbox{if } t=1 \mbox{ and } k \in \{0,1\}, \\
		0  & \mbox{if } t=1 \mbox{ and } k \notin \{0,1\}, \\
		q_{4t+k-6} & \mbox{otherwise.}
	\end{array}
\right.
\end{equation}
The remainder of the equations from the previous section is derived analogously with the number of dimensions $D$ set to $D=2$ in Eq.~\ref{eq:h_overlap} and \ref{eq:h_pair}.
The various mixer Hamiltonians also remain unchanged since we derived them in a general manner.
Note again, that the only major change between cubic and planar lattices is that a single turn is encoded with $n=4$ qubits rather than $n=6$.

\section{Results \& Discussion}
\label{sec:results_discussion}

In the current implementation the XY mixer Hamiltonian enforces the hard constraint of preserving a Hamming weight of $1$ within each $n$-tuple of qubits that encodes a turn whilst the XZ mixer also allows for $n$-tuples with Hamming weight $0$.
We will now discuss possible modifications to the mixer Hamiltonians derived in the previous section that increase the number of hard constraints while decreasing the number of soft constraints in the cost Hamiltonian.

The $H_{overlap}$ component in the current cost Hamiltonian penalizes short as well as long-range overlaps.
An overlap (self-intersection of the amino-acid chain on the lattice) is called \textit{short-range} if it occurs between the $i$-th and $(i+2)$-th residue. All other overlaps are referred to as \textit{long-range}.

\textbf{XY short-range overlap mixer.}
Short-range overlaps can be avoided in the XY formulation of the mixer Hamiltonian by introducing the controlled \texttt{SWAP} operation
\begin{equation}
\label{eq:hback_mixer}
	(\bar{q}_{t-1,\bar{k}_n},\bar{q}_{t+1,\bar{k}_n})\,\texttt{SWAP}(q_{t,k},q_{t,k_n}),
\end{equation}
which swaps turn $t$ from encoding the $k$ direction to the $k_n$ direction conditioned on the fact that the previous ($t-1$) and the next ($t+1$) turn are not encoding the $\bar{k}_n$ direction.
Writing out the product over control qubits explictly, we write
\begin{equation}
\label{eq:hback_mixer_withprod}
	\texttt{SWAP}(q_{t,k},q_{t,k_n})\, \prod^1_{j=0} \bar{q}_{t+(-1)^j,\bar{k}_n}.
\end{equation}
Using the fact that $\bar{q} = (1-q)$ and substituting $q_i = \frac{1}{2} (\mathbb{1} - Z_i)$ we obtain
\begin{equation}
\label{eq:hback_mixer_withZ}
	M_{t,k,k_n} = \frac{1}{4} \texttt{SWAP}(q_{t,k},q_{t,k_n}) \prod^1_{j=0} (\mathbb{1} + Z_{t+(-1)^j,\bar{k}_n})\, ,
\end{equation}
where $\texttt{SWAP}(q_{t,k},q_{t,k_n}) = \frac{1}{2} (X_kX_{k_n} + Y_kY_{k_n})$.
By summing over all $N-2$ turns and all directions within each turn we obtain the new mixer Hamiltonian,
\begin{equation}
\label{eq:hback_mixer_withZ}
H_{M} = \text{XY}_{short} = \sum^{N-2}_{t=1} \sum^{n-1}_{k \neq k_n} M_{t,k,k_n},
\end{equation}
which now includes the new hard constraint that solutions with short-range overlaps are not feasible and is thus referred to as XY$_{short}$.

\textbf{XZ short-range overlap mixer.}
The XZ mixer that prevents short-range overlaps is obtained by replacing the \texttt{SWAP} operation in Eq.~\ref{eq:hback_mixer_withZ} with the conditional bit flip operator such that
\begin{equation}
\label{eq:hback_mixer_withZ}
	M_{t,k} = \frac{1}{4} F_{t,k} \prod^1_{j=0} (\mathbb{1} + Z_{t+(-1)^j,\bar{k}_n}).
\end{equation}
The resulting mixer Hamiltonian will be referred to as XZ$_{short}$ and is given by,
\begin{equation}
\label{eq:hback_mixer_withZ}
H_{M} = \text{XZ}_{short} = \sum^{N-2}_{t=1} \sum^{n-1}_{k =0} M_{t,k}.
\end{equation}

In both cases, XZ and XY, moving the short range overlaps into the mixer Hamiltonian reduces the number of terms in the $H_{overlap}$ component of the cost Hamiltonian such that,
\begin{equation}
\label{eq:h_overlap}
	H_\text{olap} = \lambda_\text{olap} \sum^{N-3}_{i=0} \sum^{\lfloor\frac{(N-i-1)}{2}\rfloor}_{j=2} H_\text{olap}(i,i+2j) ,
\end{equation}
where the second sum now starts at $j=2$ instead of $j=1$.

\textbf{XY long-range overlap mixer.}
We now incorporate the entire $H_{overlap}$ into the mixer Hamiltonian.
For this we need to define the three functions,
\begin{align}
	x_a &= \begin{cases}
		0, & \text{if a=0},\\
		1 + q_0 + \sum^{a-1}_{j=2} \big( d^j_{k=0} - d^j_{k=3}\big) , & \text{otherwise}, \\
	\end{cases} \\\
	y_a &= \begin{cases}
		0, & \text{if a=0},\\
		q_1 + \sum^{a-1}_{j=2} \big( d^j_{k=1} - d^j_{k=4}\big), & \text{otherwise}, \\
	\end{cases} \\
	z_a &=  \sum^{a-1}_{j=2} \big( d^j_{k=2} - d^j_{k=5}\big).
\end{align}
which return the $x,y,z$ coordinates of the $a$-th amino acid respectively.
Using these three expressions we can define the squared lattice distance between residues $j$ and $k$:
\begin{equation}
	D_{j,k} = (x_j - x_k)^2 + (y_j - y_k)^2 + (z_j - z_k)^2.
\end{equation}
The squared lattice distance $D_{j,k}$ is zero iff residues $j$ and $k$ occupy the same lattice point.
Hence, the $D_{j,k}$ can be bounded by
\begin{equation}
	0 \leq D_{j,k} \leq (j-k)^2.
\end{equation}
This allows us to control \texttt{SWAP} operations in a way that also prevents long-range overlaps since we can simply multiply with the squared lattice distance that would be the result of turn $t$ moving into the proposed direction $k_n$.
In case of an overlap, $D_{i,t+1}=0$ and, hence, the \texttt{SWAP} operation will not be applied.
Hence, the XY mixer Hamiltonian that prevents long-range overlaps reads,
\begin{equation}
\label{eq:xy_overlap}
\begin{split}
\text{XY}_{long} = &\sum^{N-2}_{t=1} \sum^{n-1}_{k \neq k_n} M_{t,k,k_n} \\
&\prod_{i=0}^{N-5} \big(1+(D_{i,t+1}-1)[(i-t)\,\text{mod}\,2]\big).
\end{split}
\end{equation}
A convenient property of this expression is that it does not only prevent long-range overlaps but also promotes the swapping of directions if pairs of residues have a large squared lattice distance since in those cases $D_{j,k}$ is large.
This means that the transitions from extended protein chains to folded protein states are incentivized through a large multiplicative factor. We will refer to this particular definition of $H_M$ as XY$_{long}$.

\textbf{XZ long-range overlap mixer.}
Once again, we obtain the XZ mixer equivalent by replacing \texttt{SWAP} with the conditional bit flip operator in Eq.~\ref{eq:xy_overlap} which results in
\begin{equation}
\label{eq:xz_overlap}
\begin{split}
\text{XZ}_{long} = &\sum^{N-2}_{t=1} \sum^{n-1}_{k=0} M_{t,k} \\
&\prod_{i=0}^{N-5} \big(1+(D_{i,t+1}-1)[(i-t)\,\text{mod}\,2]\big).
\end{split}
\end{equation}


\begin{figure*}
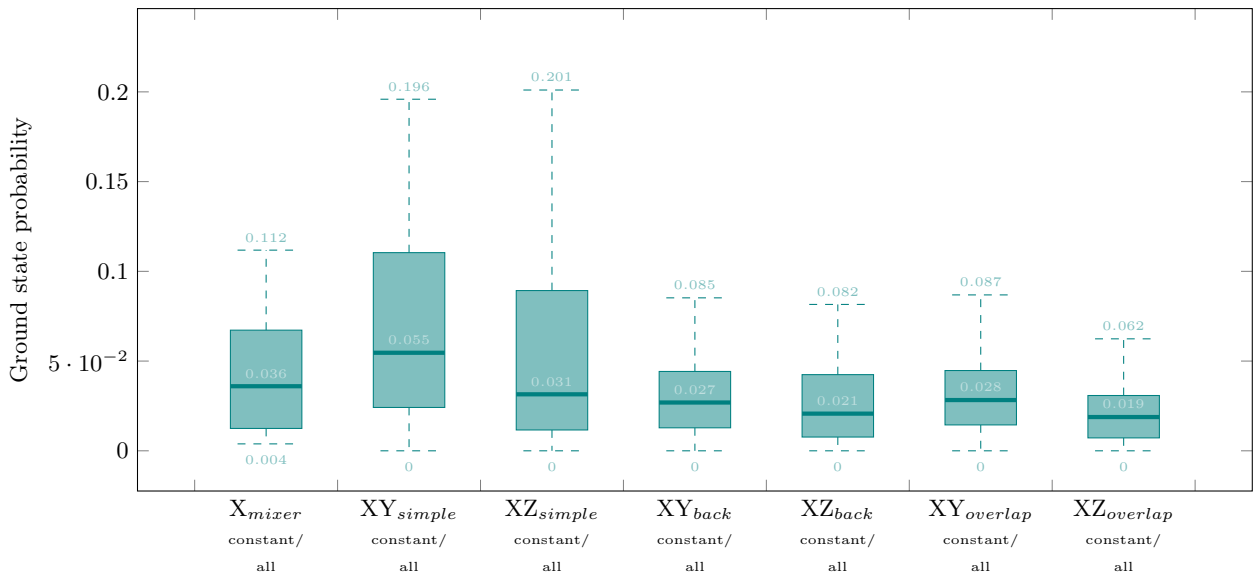


\caption{\label{plt:uniform_superpos_constant_ham}Comparison of the ground state probabilities for the folding of PSVK on a planar lattice with various mixer Hamiltonians. In this experiment the size of the cost Hamiltonian remained \textit{constant}. QAOA circuits were initialized to a uniform superposition over \textit{all} bitstrings. The classical optimizer was set to $0.001$ error tolerance and results were obtained with Rigetti's quantum computing simulator without noise. For each mixer Hamiltonian a box plot for QAOA depth $p=1$ is shown whereby the statistics were collected from $100$ runs.}
\end{figure*}


\begin{figure*}
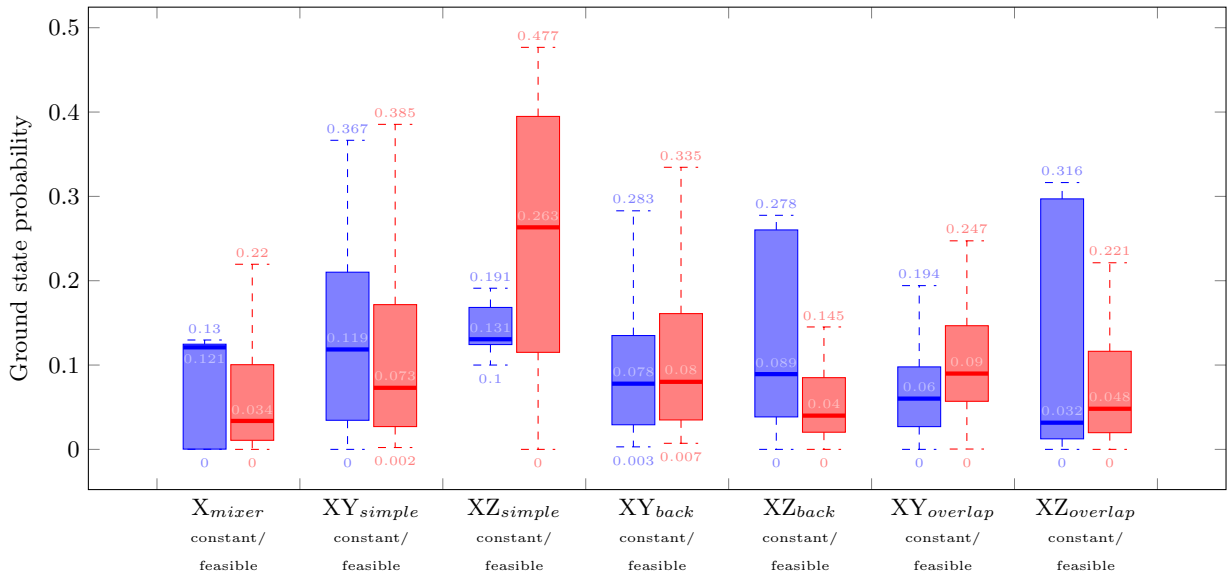


\caption{\label{plt:feasible_superpos_constant_ham}Comparison of the ground state probabilities for the folding of PSVK on a planar lattice with various mixer Hamiltonians. In this experiment the size of the cost Hamiltonian also remained \textit{constant}. Yet, QAOA circuits were initialized to a uniform superposition over all \textit{feasible} bitstrings with the initialization circuit $I$. The classical optimizer was set to $0.001$ error tolerance and results were obtained with Rigetti's quantum computing simulator without noise. For each mixer Hamiltonian a box plot for QAOA depth $p=1$ (blue) and $p=2$ (red) is shown whereby the statistics were collected from $100$ runs and $50$ runs respectively.}
\end{figure*}


\begin{figure*}
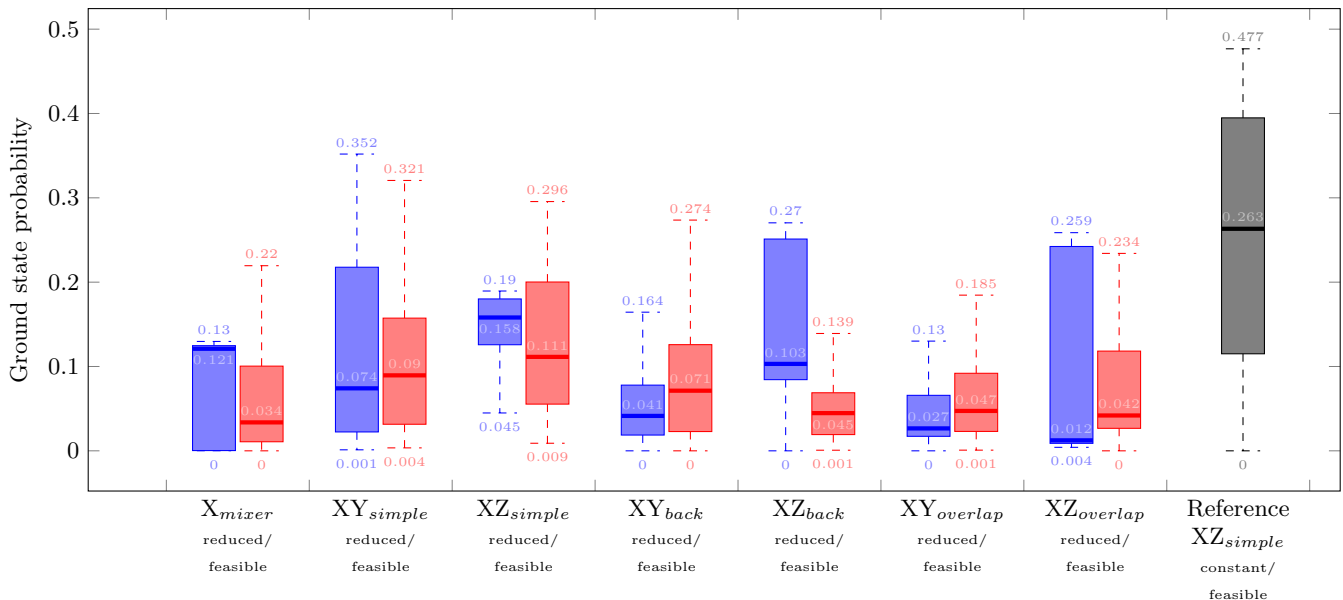


\caption{\label{plt:feasible_superpos_reduced_ham}Comparison of the ground state probabilities for the folding of PSVK on a planar lattice with various mixer Hamiltonians. In contrast to previous experiments, the size of the cost Hamiltonian was \textit{reduced} depending on the mixer Hamiltonian used. QAOA circuits were initialized to a uniform superposition over all \textit{feasible} bitstrings with the initialization circuit $I$. The classical optimizer was set to $0.001$ error tolerance and results were obtained with Rigetti's quantum computing simulator without noise. For each mixer Hamiltonian a box plot for QAOA depth $p=1$ (blue) and $p=2$ (red) is shown whereby the statistics were collected from $100$ runs and $50$ runs respectively.}
\end{figure*}

To evaluate the effectiveness of the various definitions of $H_M$, we performed extensive numerical experiments. For these experiments, we used the quantum computing simulator that is available as part of Forest, the quantum full-stack library developed by Rigetti Computing Inc. \cite{smith2016practical}.
Fig.~\ref{plt:uniform_superpos_constant_ham} shows the probabilities of obtaining the ground state solution for the two-dimensional lattice protein PSVK with the seven different definitions of the mixer Hamiltonian.
In this first experiment, the initial state was a uniform superposition over all bitstrings of length 6 and the size of the cost Hamiltonian was kept constant.
For example, this means that when using the XZ$_{back}$ mixer Hamiltonian we would not remove the short-range overlap component from the cost Hamiltonian.
This allows for an accurate baseline comparison between the pure X mixer and the other mixer Hamiltonians.
The error tolerance of the classical Nelder-Mead optimizer was chosen to be $0.001$ and the results for QAOA depth $p=1$ are shown.
The reported probabilities were averaged over $100$ runs on a hypothetical all-to-all connected quantum computer without any noise.
From the plot in Fig.~\ref{plt:uniform_superpos_constant_ham} it becomes evident that none of the other mixer Hamiltonians significantly outperforms the traditional X mixer in this setting. The median ground state probabilities range between $0.055$ for the XY$_{simple}$ and $0.019$ for the XZ$_{overlap}$. The pure X$_{mixer}$ has a median ground state probability of $0.036$ and a maximum value of $0.112$. Interestingly, when comparing maximum values the XY$_{simple}$ and XZ$_{simple}$ mixer slightly outperform the X$_{mixer}$ with maxima of $0.196$ and $0.201$ respectively. The explanation for why the new mixers are not performing well in this baseline setting is that both, XY and XZ mixers, preserve Hamming weight to various extent. \texttt{SWAP}-based mixers such as the XY mixers keep Hamming weight constant whereas XZ mixers either keep Hamming weight constant or increase/decrease the original Hamming weight by $1$. This also holds for the unfeasible states $\ket{00 \,0000}$ and $\ket{11 \,1111}$ contained in the uniform superposition over all $6$-bit strings. Yet, the XY and XZ mixers fail at converting these strings into feasible solutions whilst the X mixer can achieve this by simply flipping the respective bits.

To analyse the full potential of the new mixer Hamiltonians we repeat the previous experiment but rather than initializing the QAOA circuit with a uniform superposition over all bitstrings we initialize a superposition over \textit{all feasible} bitstrings instead. The desired initial state should look like:
\begin{equation}
\label{eq:initial_state}
\ket{\psi_i} = \ket{10\, 1000} + \ket{10 \,0100} + ... + \ket{01 \,0001},
\end{equation}
which can be achieved with a relatively shallow circuit $I$.
Using $\ket{\psi_i}$ as the initial state for the QAOA circuit we obtain the ground state probabilities shown in Fig.~\ref{plt:feasible_superpos_constant_ham}.
The Nelder-Mead error tolerance remained unchanged and the simulations were performed without noise.
We again used $100$ runs to average for QAOA depth $p=1$ but only used $50$ runs for $p=2$ due to the exponential runtime scaling of the simulation.
Note that, almost all six XY and XZ mixers dominate the performance of the traditional X mixer at QAOA depths $p=1$ (blue box plots) and $p=2$ (red box plots) in this setting.
From all mixer definitions, the XZ$_{simple}$ at depth $p=2$ clearly performs the best with a maximum ground state probability of $0.477$ and a median of $0.263$ compared to the X$_{mixer}$ with a maximum of $0.148$ and a median of $0.048$.
The fact that the maximum ground state probability three-folded clearly demonstrates the power of mixer Hamiltonians that are custom to a particular problem encoding.
Interestingly, the simple XY and XZ mixer generally perform better than the more complicated mixer Hamiltonians which take the short- and long-range interactions into account.
This can be best explained by the fact that the increased complexity leads to less 'mixing' of the amplitudes since a lot of the \texttt{SWAP} and conditional bit-flip operations are being prevented since they would otherwise lead to overlaps.
However, based on our simulation results it seems that a simple 'blind' approach to amplitude mixing has a higher chance of finding the ground state than a more controlled one.
All in all, this experiment showed that when initializing a uniform superposition over all feasible solutions any of the XY or XZ mixers (but XZ$_{back}$ with $p=2$) will almost certainly improve the ground state probabilities compared to the traditional X mixer.

Lastly, the experiment was repeated once more with an initial superposition over all feasible solutions and with cost Hamiltonians of varying size.
More specifically, depending on the mixer Hamiltonian used, parts of the cost Hamiltonian were left out since e.g. the XZ$_{overlap}$ mixer would account for short- and long-range overlaps.
This effectively reduces the size of the cost Hamiltonian whilst increasing the size of the mixer Hamiltonian.
Yet, the mixer Hamiltonian's increase is not necessarily proportional to the decrease in size of $H_C$.
Most importantly, Eq.~\ref{eq:h_olap} and \ref{equ:h_olap_ij} tend to inflate the cost Hamiltonian due to the many XNOR operations involved.
It follows that using the mixer Hamiltonian to prevent overlaps can reduce the circuit depth required to solve the optimization problem.
Furthermore, depending on the quantum chip architecture at hand this could potentially further reduce the required \texttt{CNOT} depth if the proposed XY or XZ mixer Hamiltonians lead to less topological \texttt{SWAP} operations.
Keeping the circuit depth low is of particular importance since NISQ era quantum computers, today as well as in the near future, will only allow for relatively shallow quantum circuits.
Fig.~\ref{plt:feasible_superpos_reduced_ham} shows the obtained simulation results for all seven mixer Hamiltonians with varying cost Hamiltonians. All other parameters were kept constant with respect to the two previous experiments.
Overall, the ground state probabilities clearly decreased in all XY and XZ cases compared to the previous experiment with constant cost Hamiltonian size.
The traditional X mixer, unsurprisingly, yields the same results since its use does not affect the size of the cost Hamiltonian (it does not prevent any overlaps).
The best performance is again observed with the XY$_{back}$ and XZ$_{back}$ mixer with maximum ground state probabilities of $0.352$ ($p=1$) and $0.296$ ($p=2$) respectively.
This again is an almost three-fold improvement compared to the pure X mixer underlining the importance of problem-specific mixer Hamiltonians in QAOA circuits.
In summary, the last experiment demonstrates a compromise typical for the NISQ era. Trading reduced circuit depth with slightly worse ground state probabilities might well be preferential on near-term devices since we likely will not have the ability to execute deep quantum circuits.

\begin{figure}
	\includegraphics[scale=0.3]{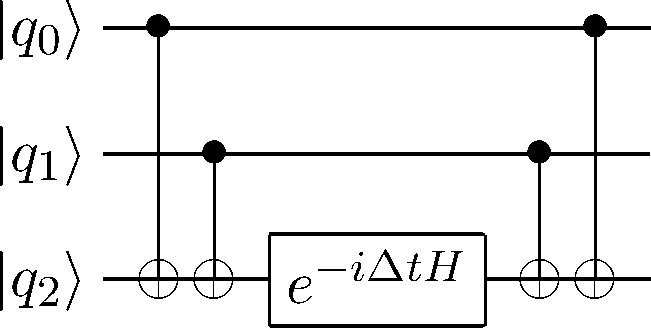}
	\caption{\label{img:zzz_circuit}Quantum circuit implementing the three-local Hamiltonian $H=\alpha (Z_0 \otimes Z_1 \otimes Z_2)$ with four \texttt{CNOT} operations and one single-qubit gate}
\end{figure}

A major issue with the one-hot encoding strategy is the locality of the cost Hamiltonian $H_C$ which is maximally local.
This means that $H_C$ is $(6N-17)$-local for a protein of size $N$ on a cubic lattice.
As shown in Fig.~\ref{img:zzz_circuit} each term in the cost Hamiltonian translates to a sequence of controlled-NOT (\texttt{CNOT}) operations and one single-qubit quantum gate.
More specifically, for a $k$-local term in $H_C$ we need to implement exactly $2k-2$ \texttt{CNOT} operations.
Additionally, some of these \texttt{CNOT} operations might not be directly implementable in the hardware due to connectivity restrictions of the qubit hardware graph.
This problem can be mitigated through the implementation of topological \texttt{SWAP} operations, which move qubits to locations where the \texttt{CNOT} operation between the desired qubits can be applied.
However, each topological \texttt{SWAP} consists of three additional \texttt{CNOT} operations which in turn further inflate the depth of the quantum circuit at hand.

Unfortunately, the quantum circuit requirements for folding the planar model protein PSVK make it impossible to directly implement the corresponding QAOA circuit on Rigetti's \textit{19Q-Acorn} processor.
For this reason, we implement a divide-and-conquer approach for the QPU implementation of this problem.
Folding PSVK on a planar lattice normally requires 6 qubits since the zeroth turn is fixed, the first turn requires 2 qubits and the last turn requires 4 qubits.
The divide-and-conquer approach consists of splitting the problem into two instances.
One in which the first turn is fixed to moving to the \texttt{RIGHT} and one in which it is fixed to moving \texttt{UP}.
In this way we reduce the problem such that the cost and mixer Hamiltonian each only consist of 4 qubits.
These two problems were separately solved on the QPU with QAOA circuits of depth $p=1$ that were classical optimized with a Nelder-Mead optimizer with an error tolerance of $0.5$. The initial state was a uniform superposition over \textit{all} bitstrings since preparing $\ket{\psi_i}$ would have pushed the quantum circuit depth beyond feasibility.
We used $10.000$ samples to approximate the expectation value of each term in the cost function and used the \texttt{XY}$_{simple}$ mixer.
In both runs we obtained a total of $2000$ samples of which $121$ yielded the correct ground state $\ket{0010}$.
Even though PSVK is a tiny 2D model protein, to our knowledge, this proof-of-principle experiment is the first time that a lattice protein was folded on a gate-based quantum computing architecture.

This work was focused on a newly introduced one-hot encoding of lattice proteins in order to investigate the use of various mixer Hamiltonians.
It remains to be investigated if more dense encodings as published in \cite{babbush2012construction}, \cite{pq_whitepaper1} or \cite{perdomo2008construction} in conjunction with a plain X mixer Hamiltonian might result in the folding of larger proteins with QAOA circuits.
The main bottlenecks are current hardware limitations, especially in terms of \texttt{CNOT} depth and qubit connectivity.
Since qubit connectivity can be mitigated using schemes such as LHZ-QAOA \cite{lhz}, the hardware metric with the biggest potential for providing performance increase is the two-qubit gate depth, determined by the qubit coherence and gate application time.
Since this has already been observed by the hardware manufacturers, it is indeed the available circuit depth which is likely to be tackled next, since the size of the state space of the hardware system in terms of qubits already surpassed the simulatable range of classical computers \cite{google}.


\section{Future Work}
\label{sec:future_work}

Future work will investigate the effects of noise on the proposed algorithm as well as explore the most efficient implementation of the variational algorithm for lattice protein folding on fixed qubit architectures with the technique introduced by \citet{farhi2017quantum}.
Furthermore, additional work is needed in optimizing the QAOA circuits for lattice protein folding with e.g. temporal planners as proposed by \citet{venturelli2018compiling} since current quantum computing hardware enforces strict upper bounds on the quantum circuit depth.
\citet{lechner2018quantum} proposed yet another way of decreasing the \texttt{CNOT} depth of QAOA circuits at the cost of using additional ancilla qubits which are used to implement all-to-all connectivity through the use of a lattice gauge model.
In recent work, \citet{mcclean2018barren} pointed out that variational algorithms should not be initialized with random circuits since it gets exponentially difficult for the classical optimizer to find a non-zero gradient when using large numbers of qubits.
Proposed solutions include structured initial guesses or pre-training of quantum circuit segments which should be investigated in the context of lattice folding.

\section{Conclusions}
\label{sec:conclusion}
This work introduced a novel encoding of the lattice protein folding problem on cubic and planar lattices using a one-hot encoding of turns on the employed lattice.
The number of required qubits in this new encoding scales linearly with the size of the protein.
The new encoding was designed such that the protein folding problem can be separated into hard and soft constraints in order to make use of the fact that universal gate-based quantum computers enable us to freely choose a mixer Hamiltonian for the QAOA circuits.
We showed how this approach can be used to incorporate various parts of the cost Hamiltonian into the mixer Hamiltonian and derived six novel problem-specific mixer Hamiltonians.
Using extensive numerical simulations, we demonstrated that the use of these problem-specific mixer Hamiltonians as well as initial states can lead to a three-fold increase in the probability of obtaining the correct protein fold.
Across all experiments, the XY$_{back}$ and XZ$_{back}$ mixer Hamiltonians performed the best with a maximum ground state probability of $0.477$.
Additionally, we performed a proof-of-concept QPU implementation.
Using a divide-and-conquer approach together with the XY$_{back}$ mixer Hamiltonian we successfully folded the mini protein PSVK on a planar lattice with a QAOA circuit on Rigetti's 19Q-Acorn processor.
With respect to near-term NISQ computers, the new encoding as well as the new mixer Hamiltonians might not only improve the solution quality but also be beneficial in terms of practicality depending on the chip topology and the available universal gate set.

\subsection*{Acknowledgments}
Special thanks to Reed McBride, Ryan Karle, Matt Harrigan and Nima Alidoust at Rigetti Computing Inc. for continued support, fruitful discussions and access to the 19Q-Acorn quantum processor.
We are grateful for the support of the Rotman School of Business Creative Destruction Lab, especially Daniel Mulet, Hassan Bhatti and Khalid Kurji.
We acknowledge financial support from Data Collective, Spectrum28, Bloomberg Beta and private investors.


\bibliographystyle{apalike}
\bibliography{qaoa_lattice_folding}

\end{document}